\documentclass[aps,reprint,longbibliography,superscriptaddress]{revtex4-1}
\usepackage{amsmath}
\usepackage{amssymb}
\usepackage{bm}
\usepackage{epsfig}
\usepackage{graphicx}
\usepackage[dvipsnames]{xcolor}
\usepackage{color}
\usepackage[unicode=true,colorlinks=true,citecolor=blue,urlcolor=blue]{hyperref}
\usepackage[T2A]{fontenc}
\usepackage[cp1251]{inputenc}
\usepackage[english]{babel}
\usepackage{braket}
\usepackage{ulem}
\usepackage{physics}

\graphicspath{{img/}}

\newcommand{\e}{\mathrm{e}}

\renewcommand{\i}{{\rm i}}

\renewcommand{\emph}{\textit}

\renewcommand{\braket}[1]{\left\langle #1 \right\rangle}

\begin{document}

\title{Optical measurement of electron spins in quantum dots: Quantum Zeno effects}
% \title{Optical measurements of the localized electron spins: Quantum Zeno effect\addDima{s}}

\author{N.~V.~Leppenen}
\email{leppenen@mail.ioffe.ru}

\author{D.~S.~Smirnov}

\affiliation{Ioffe Institute, 194021 St. Petersburg, Russia}

%\commentDima{I suggest to provide your email instead of mine.}

%\date{\today}

\begin{abstract}
  We describe the effects of the quantum back action under continuous optical measurement of electron spins in quantum dots. We consider the system excitation by elliptically polarized light close to the trion resonance, which allows for the simultaneous spin orientation and measurement. We microscopically demonstrate that the nuclei-induced spin relaxation can be both suppressed and accelerated by the continuous spin measurement due to the quantum Zeno and anti-Zeno effects, respectively. Our theoretical predictions can be directly compared with the future experimental results and straightforwardly generalized for the pump-probe experiments.
\end{abstract}

\maketitle

\section{Introduction}

The quantum Zeno effect is the well known phenomenon of freezing of the quantum dynamics of the systems under frequent measurements. This resembles the arrow paradox formulated by Zeno of Elea 25 centuries ago~\cite{doi:10.1063/1.523304}. The first prediction of the increase of the decay time of a quasi-stationary state under frequent perturbations was made by Khalfin in the late~1950s~\cite{khalfin1958contribution,khalfin_zenos_1990}, and the most popular experimental demonstration of this effect was made for a transition between two ${}^9$Be$_+$ ground-state hyperfine levels~\cite{itano_quantum_1990}. It is somewhat less known that the system measurements can also accelerate its relaxation, which is called an anti-Zeno effect~\cite{kaulakys_quantum_1997,PhysRevLett.86.2699,PhysRevLett.97.130402}.%,facchi_2008}.%bednorz_nonclassical_2012}.

%Despite the ambiguous position of the fundamental value of this effect~\cite{petrosky_quantum_1990, ballentine_comment_1991, itano_perspectives_2009},
Nowadays, both quantum Zeno and anti-Zeno effects are considered as an important tool for the quantum computing~\cite{paz-silva_zeno_2012,hacohen-gourgy_incoherent_2018}. On the one hand, it has to be taken into account during the qubit measurement, and on the other hand, it allows one to increase the relaxation time of the qubits. This leads to the growing interest in the investigation of the quantum Zeno effects for many systems promising for the quantum computations: cold atoms~\cite{chen_quantum_2021,PhysRevA.89.043614}, trapped ions~\cite{leibfried_quantum_2003}, quantum cavities and waveguides~\cite{PhysRevB.80.155326,nutz_stabilization_2019,luo_high-precision_2020,Zeno_PRB,das_interplay_2021}, NV centers in diamond~\cite{doi:10.1126/science.1131871,cujia2019tracking,pfender2019high}, and nuclear spins~\cite{klauser_nuclear_2008,PhysRevLett.125.047701,doi:10.1126/sciadv.aba3442,dasari2021antizeno}.

In the same time, the electron spins in quantum dots are believed to be one of the most prominent candidates for scalable quantum computations~\cite{petta_coherent_2005,huang_fidelity_2019,chatterjee_semiconductor_2021}. However, the quantum measurement back action is essentially unexplored experimentally and even theoretically for quantum dots. In this paper we fill this gap and develop a microscopic theory of the quantum Zeno and anti-Zeno effects for electron spins in the quantum dots resonantly measured by the laser light. We suggest the measurement of the steady state spin polarization under continuous excitation of the dots by elliptically polarized light as the simplest experiment that allows one to demonstrate the quantum back action.
  
%For example, single spins in the quantum dot micropillar cavities allow one to generate large photon cluster states~\cite{lindner_proposal_2009}. 

%The control and measurement of the single electron qubits by electromagnetic field unavoidably leads to the quantum Zeno effect in such systems. In the paper~\cite{Zeno_PRB} we have shown, that the quantum Zeno effect caused by continuous wave measurement change the spin dynamics in the quantum-dot-cavity system in the external magnetic field, which is applied to such systems in order to generate maximally entangled states~\cite{lindner_proposal_2009}.

The quantum Zeno effect requires slow and non-Markovian spin relaxation. In quantum dots this kind of relaxation is typical in small magnetic fields because of the hyperfine interaction with the slowly varying nuclear spin environment~\cite{book_Glazov,noise-CPT}. We find that the resonant optical system excitation can lead to both suppression and acceleration of the nuclei-induced spin relaxation due to the quantum Zeno and anti-Zeno effects.

%However, the main decoherence mechanism of the localized electron spin in the quantum dot is the interaction with nuclear field. Experimental and theoretical research of the electron-nuclei interaction are now actively conducted~\cite{chekhovich_measurement_2017,
%cujia_tracking_2019,PRC_General}. In this paper we study the effect  of the measurement back-action in the system with localized electrons spin under the illumination by the continuous wave elliptically polarized light with the account of the hyperfine interaction. We will show that the measurement changes the picture of the spin dynamics in the Overhauser field~\cite{book_Glazov} and the quantum Zeno and anti-Zeno effects take place.
%Interestingly, in the paper~\cite{klauser_nuclear_2008} the measurement of the nuclear spins was proposed to conserve and manipulate the Overhauser field.

%\commentDima{There were a lot of references about nuclear spins. I suggest to put them somewhere in the next section.}

The paper is organized as follows: In the next section we formulate the model of the system under study. In Sec.~\ref{sec:Th} we derive the kinetic equation for the electron spin dynamics in the quantum dots. Then in Sec.~\ref{sec:Res} we present its steady state solution and discuss the quantum Zeno and anti-Zeno regimes in detail. In particular, we find an analytical solution of the kinetic equation in these regimes. In Sec.~\ref{sec:Disc} we discuss the generalization of our results to the pulsed spin measurements. The results of the work are summarized in Sec.~\ref{sec:Concl}.

\begin{figure}[tb]
	\includegraphics[width = 0.8\linewidth]{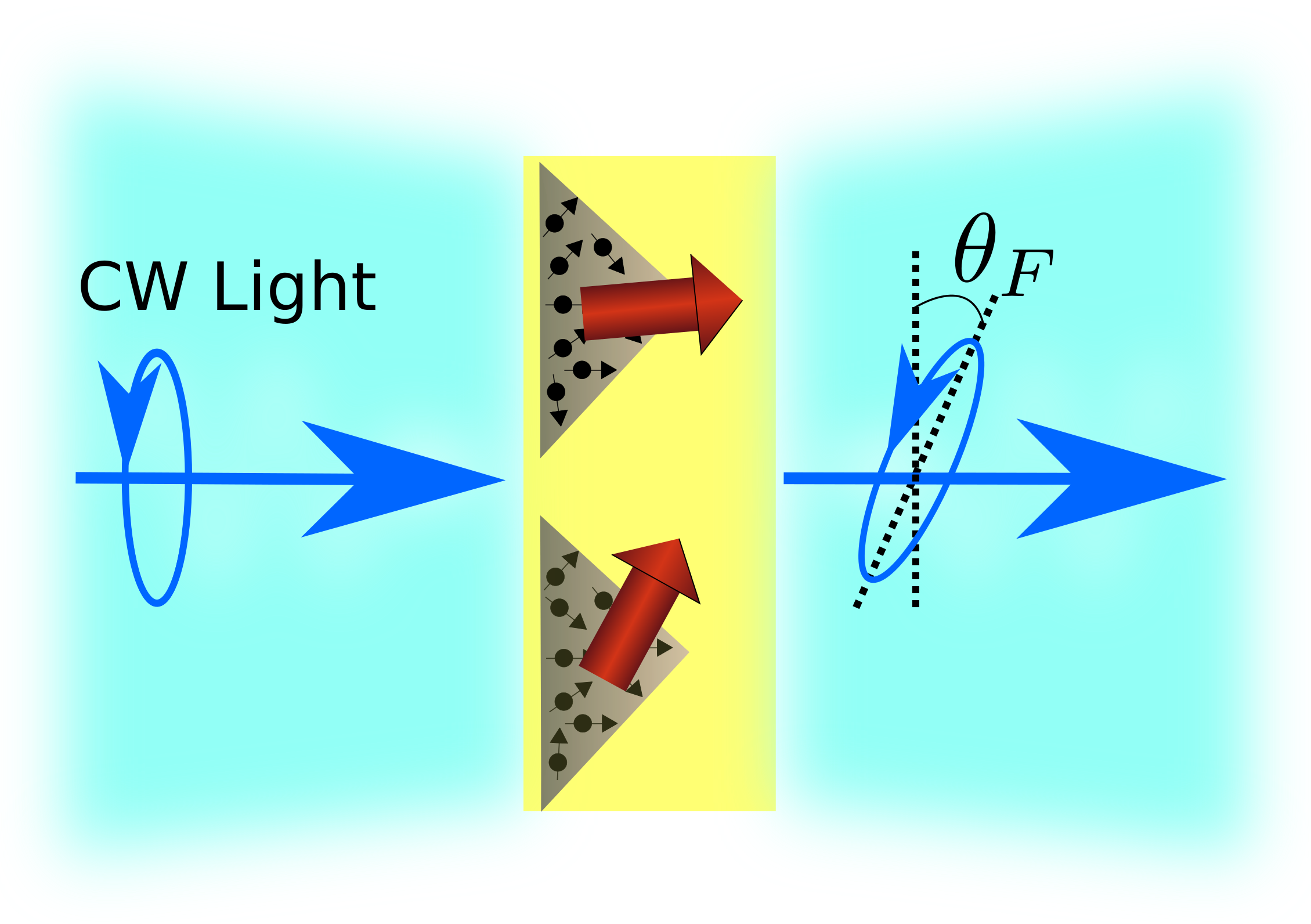} 
	\caption{Sketch of the ensemble of quantum dots (gray triangles) with the localized electron spins (red arrows) interacted with randomly oriented nuclear spins (black arrows) under illumination by CW elliptically polarized light and the measurement of the spin induced Faraday rotation angle $\theta_F$. 
%	\commentDima{Make the blue background lighter and add arrows for the elliptical polarization.}
}
	\label{fig:system_sketch}
\end{figure}

\section{Model}\label{sec:Mod}
We consider an ensemble of identical singly charged quantum dots under continuous illumination by elliptically polarized light, as shown in Fig.~\ref{fig:system_sketch}. The circular component of the light orients electron spins in the quantum dots, while the linear component allows for the measurement of the spin induced Faraday rotation~\cite{PhysRevLett.127.157401}. We assume the frequency of the light $\omega$ to be close to the singlet trion resonance frequency $\omega_0$. The singlet trion consists of two electrons with antiparallel spins and a heavy hole with the spin projection $J_z=\pm3/2$ on the structure growth axis $z$.

%illumination of elliptically polarized continuous wave (CW) light. The sketch of the system is shown in  Figure~\ref{fig:system_sketch}.

\begin{figure}[t]
	\includegraphics[width=0.8\linewidth]{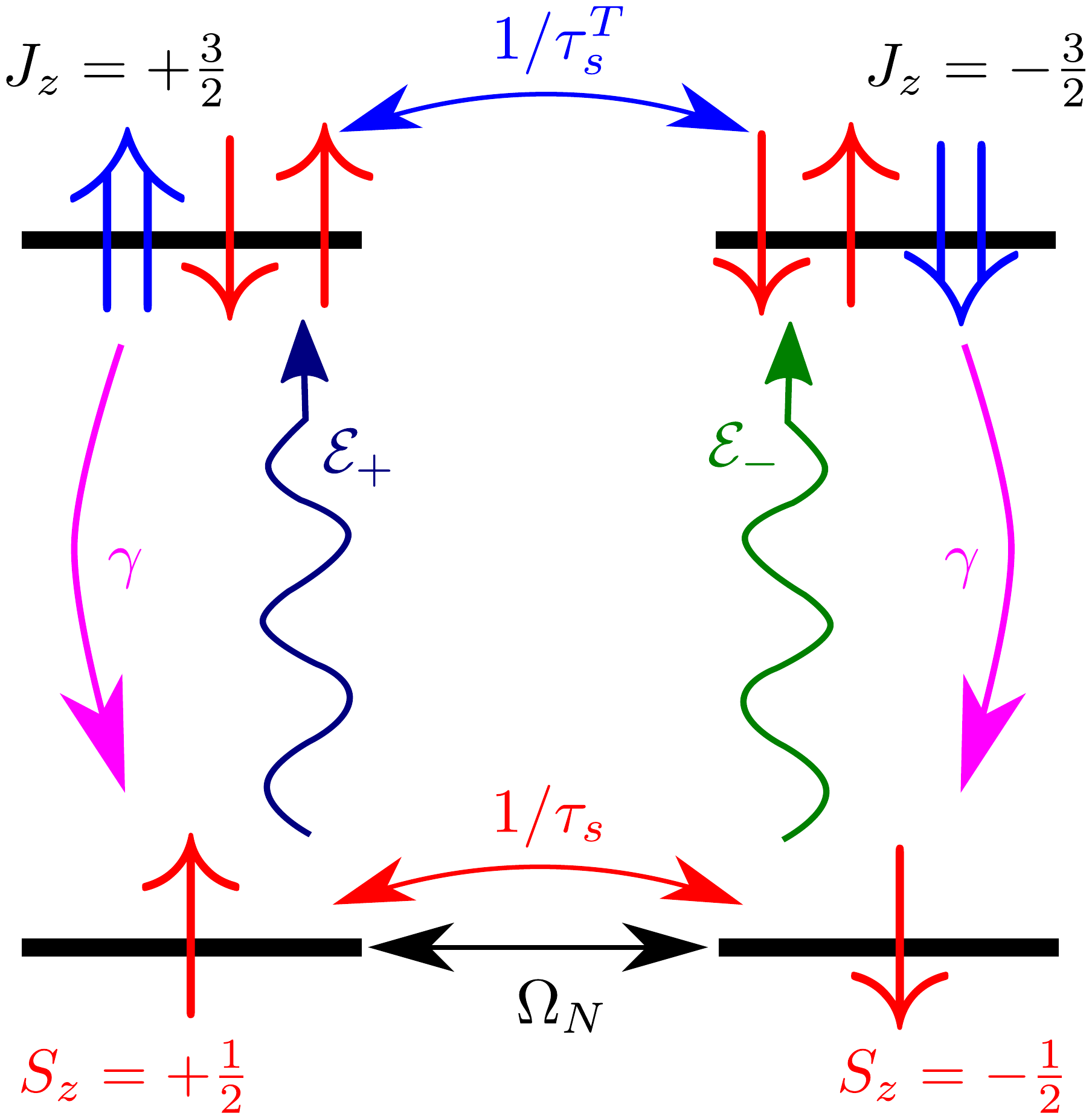} 
	\caption{Energy levels of the quantum dot and transitions between them. The two ground states are characterized by the electron spin $S_z=\pm1/2$, and the two singlet trion states by the hole spin $J_z=\pm3/2$. Absorption of $\sigma^+$ and $\sigma^-$ photons is shown by the blue and green wavy vertical arrows. The magenta arrows show the trion recombination. The blue and red double arrows show the electron and trion spin relaxations, respectively. Finally, the black double arrow shows the electron spin precession in the random nuclear field.
	}
	\label{fig:en_scheme}
\end{figure}

The Hamiltonian of the system accounting for the hyperfine interaction with the host lattice nuclei %and the trion excited states
has the form~\cite{singlespin}: 
%\commentDima{Maybe reduce number of references.}
%with an energy $\hbar \omega_0$. It could be written in the form
\begin{gather}
  {\cal H}  = \frac{\hbar \bm \Omega_N}{2}\sum_{i,j = \pm 1/2} \bm \sigma_{ij} a_i^\dagger a_j+
  \sum_{i = \pm 3/2}\hbar \omega_0 a^\dagger_{i} a_{i}\\ \notag+ \hbar\qty[\mathcal{E}_+\e^{-\i\omega t} a^\dagger_{+3/2}a_{+1/2}+\mathcal{E}_-\e^{-\i\omega t}a^\dagger_{-3/2}a_{-1/2}+\text{H.c}].
\end{gather}
%\commentDima{I suggest to replace $e$ with $\e$ and $i$ with $\i$ everywhere.}
Here $a_i$ ($a_i^\dagger$) are the annihilation (creation) operators of the electron ground states with the spins $S_z = \pm 1/2$ for $i=\pm1/2$, respectively, and of the trion states with the spins $J_z = \pm 3/2$ for $i=\pm3/2$, respectively. Further, $\bm\Omega_N$ is the electron spin precession frequency related to the random Overhauser field of the nuclear spin fluctuations~\cite{book_Glazov,PRC_General}, and $\bm\sigma$ is the vector composed of the Pauli matrices. We neglect the much weaker hyperfine interaction for holes in trions. Finally, the parameters $\mathcal{E}_\pm$ are proportional to the amplitudes of the $\sigma^\pm$ polarized components of the incident light~\cite{milburn}. According to the optical selection rules, $\sigma^\pm$ polarized light couples the states $S_z=\pm1/2$ with the states $J_z=\pm3/2$, respectively, only, see Fig.~\ref{fig:en_scheme}.
%photon leads to the creation of the \addDima{trion} with $J_z = \pm 3/2$ and an electron with $S_z = \mp 1/2$, respectively~\cite{ivchenko05a}. Photoexcited electron and hole bound with the existing electron, that has the opposite spin to the photoexcited electron due to the Pauli principle, and set up the trion with an energy $\hbar \omega_0$. Finally, the hyperfine interaction with the nuclei leads to the electron spin precession in the ground state  \commentNikita{Why not in the excited state?} in the Overhauser field~\cite{book_Glazov}, with the precession frequency $\bm \Omega_N$, that in terms of the annihilation and creation operators are written using vector consist of Pauli spin matrices $\bm \sigma = (\sigma_x, \sigma_y, \sigma_z)$.

The nuclear spin dynamics is much slower than the electron one, which makes the electron spin dynamics non-Markovian, which is necessary for the quantum Zeno and anti-Zeno effects. Theoretically, one can consider the nuclear spins to be frozen and normally distributed~\cite{merkulov02,book_Glazov}. The hyperfine interaction for electrons is isotropic, the distribution function of the random nuclear field has a simple Gaussian form:
\begin{equation}\label{eq:dist_func}
	\mathcal{F}(\bm \Omega_N) = \frac{1}{(\sqrt{\pi} \delta)^3}\exp(-\frac{\Omega_N^2}{\delta^2}),
\end{equation}
where parameter $\delta$ determines the dispersion. Despite the continuous spin orientation, we neglect the dynamic nuclear spin polarization.% is the typical precession frequency in the Overhauser field.

Additionally, we take into account the incoherent processes: trion recombination with the rate $\gamma$, electron and trion spin relaxations with the times $\tau_s$ and $\tau_s^T$, respectively, which are unrelated with the hyperfine interaction. These processes are shown in Fig.~\ref{fig:en_scheme} and can be described theoretically in the density matrix formalism using the Lindblad superoperator~\cite{backaction,milburn,carmichael}
\begin{equation}
	\mathcal{L}\{\rho\} = \sum_{i} \gamma_i \qty(\mathcal{O}_i^\dagger{\cal O}_{i}\rho+\rho{\cal O}_i^\dagger {\cal O}_i-2 {\cal O}_i\rho{\cal O}_i^\dagger),
	% \hspace{-0.5 cm}\mathcal{L}\{\rho(t)\} = \sum_{i} \gamma_i \qty(\mathcal{O}_i^\dagger{\cal O}_{i} \rho(t)+\rho(t){\cal O}_i^\dagger {\cal O}_i-2 {\cal O}_i \rho(t){\cal O}_i^\dagger),
\end{equation}
where $\rho$ is the density matrix of the four level system and the eight operators $\mathcal{O}_i$ are taken from the set
\begin{align}
% \begin{multline}
  \{a^\dagger_{\pm 1/2}a_{\pm 3/2}, \; a^\dagger_{\pm 1/2}a_{\mp 1/2}/2,\; (a^\dagger_{+1/2}a_{+1/2}-a^\dagger_{-1/2}a_{-1/2})/2,  \nonumber\\
  a^\dagger_{\pm 3/2}a_{\mp 3/2}/2,\; (a^\dagger_{+3/2}a_{+3/2}-a^\dagger_{-3/2}a_{-3/2})/2 \}
  \nonumber
\end{align}
% \end{multline}
% $\{a^\dagger_{\pm 1/2}a_{\pm 3/2}, \; a^\dagger_{\pm 1/2}a_{\mp 1/2}/2,\; (a^\dagger_{+1/2}a_{+1/2}-a^\dagger_{-1/2}a_{-1/2}),\; a^\dagger_{\pm 3/2}a_{\mp 3/2}/2,\; (a^\dagger_{+3/2}a_{+3/2}-a^\dagger_{-3/2}a_{-3/2}) \}$
with the corresponding decay rates $\gamma_i$ from the set $\left\{\gamma, 1/\tau_s, 1/(2\tau_s), 1/\tau_s^T, 1/\qty(2\tau_s^T)\right\}$.
  % \[
  %   \gamma_i = \{\gamma, 1/\tau_s, 1/(2\tau_s), 1/\tau_s^T, 1/\qty(2\tau_s^T)\}.
  % \]
The total spin dynamics is described by the quantum master equation 
\begin{equation}\label{eq:mast_eq}
  \dot{\rho}(t) = \frac{\i}{\hbar}\qty[\rho(t),{\cal H}]-\mathcal{L}\{\rho(t)\},
\end{equation}
where the dot denotes the time derivative.

The main focus of our paper is on the quantum dots, however the same model describes many other systems with the localized electrons: shallow donors in bulk semiconductors~\cite{book_Glazov}, electrons localized at the imperfections of quantum wells~\cite{noise-trions}, and electrons in moir\'e potential of twisted transition metal dichalcogenides bilayers~\cite{Yue1701696,MX2_Avdeev,PhysRevB.104.L241401,brotonsgisbert2021moiretrapped}.

In the next section, we obtain the kinetic equation for the localized electron spin dynamics under weak measurements.

\section{Derivation of kinetic equation}\label{sec:Th}

% To derive the equation on the electron spin we use density matrix in the basis of the following states. They are shown by horizontal black lines in Figure~\ref{fig:en_scheme}:
% \begin{itemize}
% \item the ground state of the electron $\ket{\uparrow}$ and $\ket{\downarrow}$ with the spin projection $S_z = +\frac{1}{2}$ and $S_z = -\frac{1}{2}$, respectively;

% \item the excited states $\ket{\Uparrow}$ and $\ket{\Downarrow}$ that corresponds to the trion with the hole in trion momentum $J_z =+ \frac{3}{2}$ and $J_z = -\frac{3}{2}$, respectively.
% \end{itemize}

% The overall density matrix has the dimension $4 \times 4$.

\subsection{Absence of excitation}

It is useful for the following to consider the spin dynamics in the absence of the incident light, $\mathcal{E}_{\pm} = 0$, when the trions are not excited.
% We start with the case when there is not any illumination, that corresponds to the $\mathcal{E}_{\pm} = 0$.
% In this case, only the two ground spin states are populated and only the four density matrix elements are nonzero. From the master equation~\eqref{eq:mast_eq}, we obtain the following equations on the elements of $\rho(t)$
% \begin{subequations}\label{eq:sys_e_0}
% \begin{multline}
% 	\dot{\rho}_{\uparrow,\uparrow} = -\frac{\rho_{\uparrow,\uparrow}-\rho_{\downarrow,\downarrow}}{2\tau_s}+i\Omega_{N,x}\frac{\rho_{\uparrow,\downarrow}-\rho_{\downarrow,\uparrow}}{2}-\\-\Omega_{N,y}\frac{\rho_{\uparrow,\downarrow}+\rho_{\downarrow,\uparrow}}{2},
% \end{multline}
% \begin{equation}
% 	\dot{\rho}_{\downarrow,\downarrow} = \frac{\rho_{\uparrow,\uparrow}-\rho_{\downarrow,\downarrow}}{2\tau_s}+i\Omega_{N,x}\frac{\rho_{\uparrow,\downarrow}-\rho_{\downarrow,\uparrow}}{2}-\Omega_{N,y}\frac{\rho_{\uparrow,\downarrow}+\rho_{\downarrow,\uparrow}}{2},
% \end{equation}
% \begin{multline}\label{eq_rho_u_d_e_0}
% 	\dot{\rho}_{\uparrow,\downarrow}= -\frac{\rho_{\uparrow,\downarrow}}{\tau_s}-i\Omega_{N,z}\rho_{\uparrow,\downarrow}+\\+i(\Omega_{N,x}-i\Omega_{N,y})\frac{\rho_{\uparrow,\uparrow}-\rho_{\downarrow,\downarrow}}{2},
% \end{multline}
% \end{subequations}
% and the equation on $\dot{\rho}_{\downarrow,\uparrow}$ could be obtained from the complex conjugation of the expression~\eqref{eq_rho_u_d_e_0}.
The electron spin is given by the trace of Pauli matrices with the density matrix:
\begin{equation}\label{eq:spin_trace}
  \bm S(t) \equiv \Tr[\rho(t)\bm \sigma/2].
\end{equation}
From Eq.~\eqref{eq:mast_eq} in this limit we obtain the standard Bloch equation~\cite{Werner1977,book_Glazov}
\begin{equation}\label{eq:spin_zero}
  \dot{\bm S}(t) = \bm \Omega_N \times \bm S(t)-\frac{\bm S(t)}{\tau_s}.
\end{equation}
Thus, the electron spin dynamics in the absence of the excitation represents precession in the random nuclear field $\bm\Omega_N$ and relaxation with the time $\tau_s$. In what follows we consider the experimentally relevant limit of $\tau_s\gg 1/\delta$. %~\cite{PRC_General}.

\begin{figure}[t]
	\includegraphics[width=0.95\linewidth]{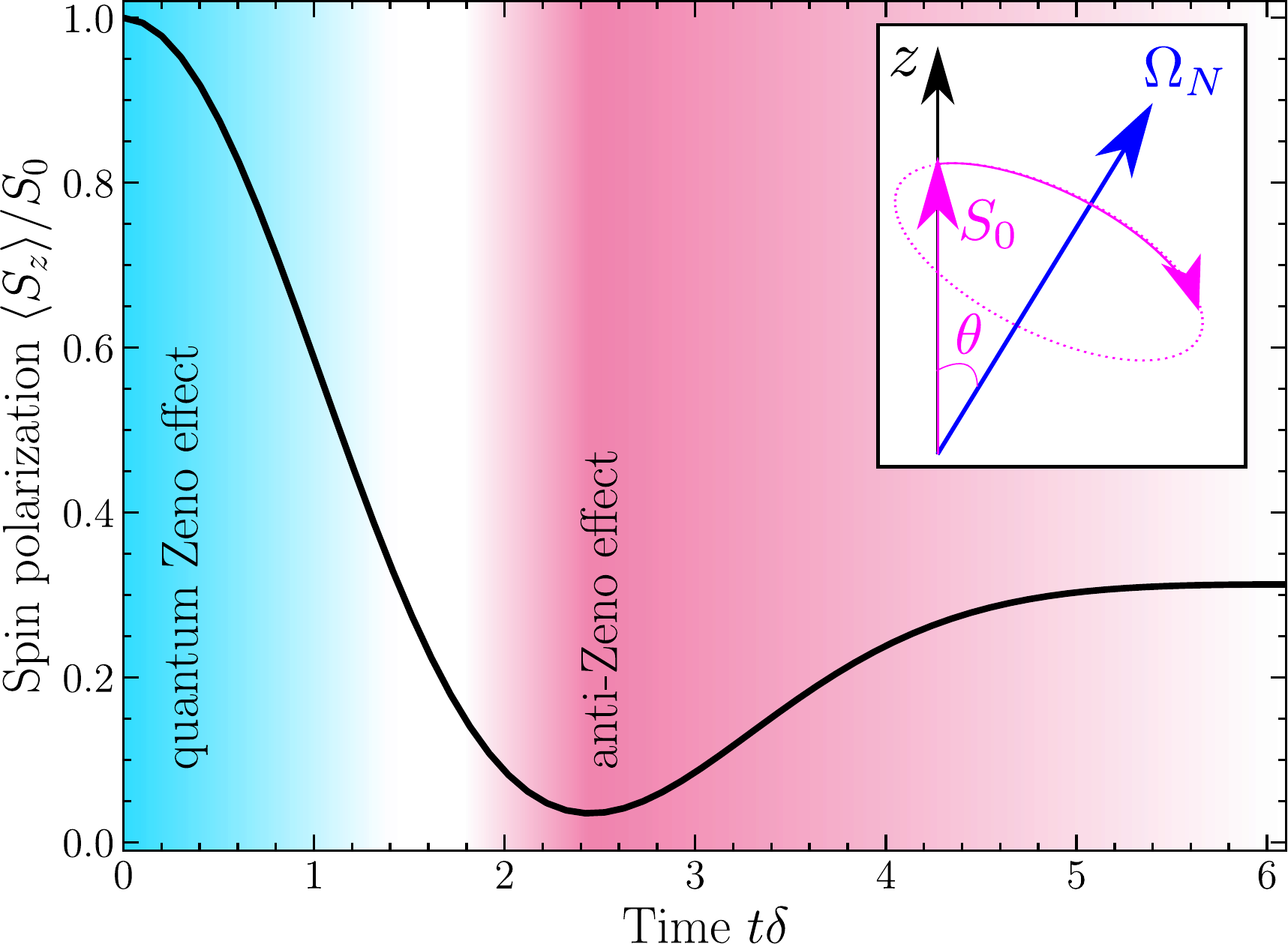} 
	\caption{Decay of the spin polarization in time due to the nuclear spin fluctuations, calculated after Eq.~\eqref{eq:spin_time} for $\tau_s\delta = 100$. The inset illustrates the electron spin precession in the random nuclear field with the frequency $\bm\Omega_N$. The light blue and coral colors show the normalized repetition periods $T_R\delta$, for which the quantum Zeno and anti-Zeno effects, respectively, are expected for the strong pulsed measurements. 
%	\commentDima{I suggest to make the white stripe between them larger.} \commentDima{I suggest to make lines in the inset thicker and labels larger.} \commentDima{Add quantum before ``Zeno''?}
}
	\label{fig:deph_time}
\end{figure}

For the given initial electron spin orientation along the $z$ axis, $\bm S(0) =S_0 \bm e_z$, the solution of the Bloch equation averaged over the nuclear field distribution function~\eqref{eq:dist_func} gives~\cite{book_Glazov}
  \begin{equation}\label{eq:spin_time}
    \expval{S_z(t)} = \left[\frac{S_0}{3}+\frac{2S_0}{3}\qty(1-\frac{t^2\delta^2}{2})\exp(-\frac{t^2 \delta^2}{4})\right]\e^{-t/\tau_s}. 
  \end{equation}
This dependence is shown by the black curve in Fig.~\ref{fig:deph_time}. The electron spin precession in the random nuclear field leads to the loss of the two thirds of the spin polarization during the time $t\sim 1/\delta$. As a result, for $1/\delta\ll t\ll\tau_s$ the electron spin polarization equals to
\begin{equation}\label{eq:s0over3}
  \expval{S_z(t)} = S_0/3.
\end{equation}
This result will be used below as a reference.
%One can see that after the time $t\sim 5/\delta$ electron spin loses 2/3 of the initial polarization.

\subsection{Weak excitation}

Now let us describe the effect of the weak system excitation by the continuous light on the spin dynamics. For its description we use the perturbation theory in small parameters ${\cal E}_{\pm}/\gamma$, which we assume to be real. We consider also the realistic limit of
\begin{equation}
  \gamma\gg 1/\tau_s^T\gg\delta\gg1/\tau_s,
\end{equation}
when the trion recombination is the fastest process in the system (typically, $\gamma\sim1$~ns$^{-1}$) and the spin relaxation in the excited trion state is faster than in that in the ground state.

In zeroth order of the perturbation theory, the four density matrix elements $\rho_{\uparrow/\downarrow,\uparrow/\downarrow}$ between the two ground states are nonzero. In the absence of excitation the equations for them yield Eq.~\eqref{eq:spin_zero}.

In the first order of the perturbation theory, there are 8 equations for the density matrix elements between the two ground and the two excited quantum dots states.
From~\eqref{eq:mast_eq} we obtain
\begin{subequations}\label{eq:first_order}
\begin{equation}\label{eq:first_order_1}
	\dot{\rho}_{\uparrow,\Uparrow} = -\qty(\frac{3}{8\tau_s^T}+\gamma-\i\omega_0)\rho_{\uparrow,\Uparrow}+\i{\cal E}_+ \rho_{\uparrow,\uparrow},
\end{equation}
\begin{equation}\label{eq:first_order_2}
	\dot{\rho}_{\uparrow,\Downarrow} = -\qty(\frac{3}{8\tau_s^T}+\gamma-\i\omega_0)\rho_{\uparrow,\Downarrow}+\i{\cal E}_- \rho_{\uparrow,\downarrow}.
\end{equation}
\end{subequations}
The equations for $\rho_{\downarrow,\Uparrow}$ and $\rho_{\downarrow,\Downarrow}$ can be obtained from~\eqref{eq:first_order_1} and~\eqref{eq:first_order_2}, respectively, by flipping the spins and exchanging ${\cal E}_+$ and ${\cal E}_-$. Four more equations for the off diagonal density matrix elements can be obtained from the hermiticity of the density matrix. These equations describe the trion excitation, recombination and spin relaxation. In the adiabatic approximation valid when the trion recombination time $1/\gamma$ is much shorter than the time scale of the spin dynamics in the ground state these density matrix elements oscillate as $e^{\pm\i\omega t}$. This allows us to find these 8 density matrix elements as functions of the ground state matrix elements~$\rho_{\uparrow/\downarrow,\uparrow/\downarrow}$.% Then equations on $\rho_{\Uparrow,\uparrow}$, $\rho_{\Uparrow,\downarrow}$, $\rho_{\Downarrow, \uparrow}$ and $\rho_{\Downarrow,\downarrow}$ are obtained from complex conjugation of the equation for the corresponding elements.

Further, in the second order, there are 4 density matrix elements between the two excited trion states. Two of them read
% Remaining 4 elements are not zero in the second order of ${\cal E}_{\pm}/\gamma$ and could be found from the equations 
\begin{subequations}\label{eq:second_order}
  \begin{equation}
    \dot{\rho}_{\Uparrow,\Uparrow} = -2\gamma \rho_{\Uparrow,\Uparrow}-\frac{\rho_{\Uparrow,\Uparrow}-\rho_{\Downarrow,\Downarrow}}{2\tau_s^T}+\i{\cal E}_+(\rho_{\Uparrow,\uparrow}-\rho_{\uparrow,\Uparrow}),
  \end{equation}
  % \begin{equation}
  % \dot{\rho}_{\Downarrow,\Downarrow} =-2\gamma \rho_{\Uparrow,\Uparrow}-\frac{\rho_{\Uparrow,\Uparrow}+\rho_{\Downarrow,\Downarrow}}{2\tau_s^T}+i{\cal E}_-(\rho_{\downarrow,\Downarrow}-\rho_{\Downarrow,\downarrow}),
  % \end{equation}
  \begin{equation}
    \dot{\rho}_{\Uparrow,\Downarrow} = -\qty(2\gamma+\frac{1}{\tau_s^T})\rho_{\Uparrow,\Downarrow}+\i({\cal E}_-\rho_{\Uparrow,\downarrow}-{\cal E}_+\rho_{\uparrow,\Downarrow}),
  \end{equation}
\end{subequations}
and the other two equations can be obtained by the spin flips and exchange of ${\cal E}_+$ and ${\cal E}_-$. Here the time derivatives should be set to zero, and using Eqs.~\eqref{eq:first_order} we find these matrix elements as functions of the density matrix of the ground state.

Finally, the equations of motion for the density matrix elements between the ground states together in zeroth and second orders read
  \begin{widetext}
  \begin{subequations}\label{eq:rho_all}
    \begin{equation}
      \dot{\rho}_{\uparrow,\uparrow} =  -\frac{\rho_{\uparrow,\uparrow}-\rho_{\downarrow,\downarrow}}{2\tau_s}+\i\Omega_{N,x}\frac{\rho_{\uparrow,\downarrow}-\rho_{\downarrow,\uparrow}}{2}-\Omega_{N,y}\frac{\rho_{\uparrow,\downarrow}+\rho_{\downarrow,\uparrow}}{2}+2\gamma \rho_{\Uparrow,\Uparrow}+\i{\cal E}_+(\rho_{\uparrow,\Uparrow}-\rho_{\Uparrow,\uparrow}),
    \end{equation}
    \begin{equation}
      \dot{\rho}_{\downarrow,\downarrow} = \frac{\rho_{\uparrow,\uparrow}-\rho_{\downarrow,\downarrow}}{2\tau_s}-\i\Omega_{N,x}\frac{\rho_{\uparrow,\downarrow}-\rho_{\downarrow,\uparrow}}{2}+\Omega_{N,y}\frac{\rho_{\uparrow,\downarrow}+\rho_{\downarrow,\uparrow}}{2}+2\gamma \rho_{\Downarrow,\Downarrow}+\i{\cal E}_- (\rho_{\downarrow,\Downarrow}-\rho_{\Downarrow,\downarrow}),
    \end{equation}
    \begin{equation}
      \label{eq:last}
      \dot{\rho}_{\uparrow,\downarrow} = -\frac{\rho_{\uparrow,\downarrow}}{\tau_s}-\i\Omega_{N,z}\rho_{\uparrow,\downarrow}+i(\Omega_{N,x}-\i\Omega_{N,y})\frac{\rho_{\uparrow,\uparrow}-\rho_{\downarrow,\downarrow}}{2}+\i({\cal E}_-\rho_{\uparrow,\Downarrow}-{\cal E}_+\rho_{\Uparrow,\downarrow}),
    \end{equation}
\end{subequations}
\end{widetext}
% \begin{subequations}
% 	\begin{equation}
% 		\dot{\rho}_{\uparrow,\uparrow}^{(2)} =  2\gamma \rho_{\Uparrow,\Uparrow}+i{\cal E}_+(\rho_{\Uparrow,\uparrow}-\rho_{\uparrow,\Uparrow}),
% 	\end{equation}
% 	\begin{equation}
% 		\dot{\rho}_{\downarrow,\downarrow}^{(2)} = 2\gamma \rho_{\Downarrow,\Downarrow}+i{\cal E}_- (\rho_{\Downarrow,\downarrow}-\rho_{\downarrow,\Downarrow}),
% 	\end{equation}
% 	\begin{equation}
% 		\dot{\rho}_{\uparrow,\downarrow}^{(2)} = i({\cal E}_+\rho_{\Uparrow,\downarrow}-{\cal E}_-\rho_{\uparrow,\Downarrow}).
% 	\end{equation}
% \end{subequations}
and equation for $\rho_{\downarrow,\uparrow}$ can be obtained from Eq.~\eqref{eq:last} by the complex conjugation.

After substitution of the solution of Eqs.~\eqref{eq:first_order} and~\eqref{eq:second_order} in Eqs.~\eqref{eq:rho_all} we obtain the equation for the spin dynamics in the second order in $\mathcal E_\pm/\gamma$. It can be compactly written as
\begin{multline}\label{eq:spin_kin_main}
  \dot{\bm S}(t) =\bm \Omega_N \times \bm S(t)-\frac{\bm S(t)}{\tau_s}+\qty(g-\frac{S_z(t)}{\tilde{\tau}})\bm e_z-\\-2\lambda(S_x(t) \bm e_x +S_y(t) \bm e_y)+\tilde{\Omega} \bm e_z \times \bm S(t),
\end{multline}
where the parameters $g$, $\tilde\tau$, $\lambda$, and $\tilde\Omega$ represent, respectively, the spin generation rate, additional spin relaxation time, measurement strength, and additional spin precession frequency. They are all proportional to the second power of the incident light amplitude, $\mathcal E_\pm$.

The spin generation rate is given by
  \begin{equation}\label{eq:for_g}
    g = \frac{{\cal E}_-^2-{\cal E}_+^2}{4\tau_s^T[(\omega-\omega_0)^2+\gamma^2]}.
  \end{equation}
  Naturally, it is proportional to the circular polarization of the incident light and is the largest at the trion resonance frequency, $\omega=\omega_0$. It is also proportional to the trion spin relaxation rate $1/\tau_s^T$, because in the absence of the spin relaxation, the trion excitation and recombination does not produce the spin polarization~\cite{PhysRevB.98.125306,PhysRevB.98.121304}. 

Also, excitation by circularly polarized light produces effective magnetic field along the optical axis
  \begin{equation}
    \tilde{\Omega} = \frac{({\cal E}_+^2-{\cal E}_-^2)(\omega-\omega_0)}{\gamma^2+(\omega-\omega_0)^2}.
  \end{equation}
due to the dynamic Stark effect~\cite{yugova_pump-probe_2009,OpticalField}. This field is an odd function of the laser detuning from the trion resonance, $\omega-\omega_0$. 

Further, the trion excitation accelerates the longitudinal spin relaxation, which is described by the rate
  \begin{equation}\label{eq:for_lambda}
    \frac{1}{\tilde{\tau}} = \frac{{\cal E}_+^2+{\cal E}_-^2}{2\tau_s^T\qty[(\omega-\omega_0)^2+\gamma^2]}.
  \end{equation}
  Clearly, it is proportional to the trion population and the trion spin relaxation rate, $1/\tau_s^T$.

Finally and most importantly, the electron spin measurement by light leads to the quantum back action, which suppresses the coherence between the eigenstates of the observable $\sigma_z$~\cite{caves_quantum-mechanical_1987,Gross_2018,bednorz_nonclassical_2012}. This effect is often described phenomenologically~\cite{presilla199695}, while in our model we obtain the microscopic expression for the measurement strength
\begin{equation}
  \lambda = \frac{({\cal E}_+^2+{\cal E}_-^2)\gamma}{2[(\omega-\omega_0)^2+\gamma^2]}.
\end{equation}
From Eq.~\eqref{eq:spin_kin_main} one can see that the corresponding term indeed suppresses the transverse spin components $S_x$ and $S_y$ in agreement with the general principles of the quantum mechanics. Microscopically, it is caused by the trion excitation and recombination, which mainly conserves the longitudinal spin component, but not the transverse ones~\cite{zhukov_optical_2010}.

The occupancies of the trion states read
\begin{equation}
	\rho_{\Uparrow/\Downarrow,\Uparrow/\Downarrow} = \frac{1}{2}\frac{{\cal E}^2_{+/-}}{(\omega-\omega_0)^2+\gamma^2}.
\end{equation}
So the measurement strength equals to the total trion recombination rate~\cite{noise-trions,Nonresonant_nonequilibrium}
\begin{equation}
  \lambda = \gamma(\rho_{\Uparrow,\Uparrow}+\rho_{\Downarrow,\Downarrow}).
\end{equation}
For comparison, in the quantum dot-micropillar system in the strong coupling regime, the measurement strength equals to the sum of the trion decay rate and the photon escape rate from the microcavity~\cite{Zeno_PRB}. This happens because in the steady state the total recombination rate of the excited states equals to the generation rate from the ground state, which unavoidably leads to the loss of the transverse spin components.

We note that the kinetic equation~\eqref{eq:spin_kin_main} allows one to describe both the steady state and the electron spin dynamics.

\section{Results}
\label{sec:Res}
To describe the quantum Zeno and anti-Zeno effects we consider the steady state of the system. 

For simplicity we consider the resonant excitation, $\omega=\omega_0$, when the optical field $\tilde\Omega$ vanishes. Also we note that the additional spin relaxation rate $1/\tilde\tau$ can be neglected in comparison with $1/\tau_s$. In this case the kinetic equation~\eqref{eq:spin_kin_main} simplifies to
\begin{multline}\label{eq:spin_kin_res}
	\dot{\bm S}(t) = \bm \Omega_N\times \bm S(t)-\frac{\bm S(t)}{\tau_s}+g\bm e_z -\\-2\lambda (S_x(t)\bm e_x+S_y(t) \bm e_y).
\end{multline}

If the hyperfine interaction were absent, $\bm\Omega_N=0$, the steady state spin polarization would be $S_0=g\tau_s$ along the $z$ axis. The ratio of steady state spin polarization $\expval{S_z}$ and $S_0$ reveals the role of the hyperfine interaction in the electron spin relaxation.

Generally, the steady state solution of Eq.~\eqref{eq:spin_kin_main} reads
\begin{equation}\label{eq:Sz_no_average}
	S_z = \frac{S_0[1+4\lambda \tau_s+(4\lambda^2+\Omega_{N}^2\cos^2(\theta))\tau_s^2]}{1+4\lambda \tau_s+(4\lambda^2+\Omega_N^2)\tau_s^2+2\lambda \Omega_{N}^2\sin^2(\theta)\tau_s^3},
\end{equation}
where $\theta$ is the angle between $\bm \Omega_N$ and the $z$ axis, see the inset in Fig.~\ref{fig:deph_time}. This expression should be averaged over the distribution function~\eqref{eq:dist_func} to obtain the average spin polarization $\expval{S_z}$.

% In the case when $\lambda = 0$ the steady state solution of the equation~\eqref{eq:spin_kin_res} leads to
% \begin{equation}
% 	S_z = \frac{g \tau_s (1+\Omega_{N,z}^2\tau_s^2)}{1+\Omega_N^2\tau_s^2} = S_0\frac{1+\Omega_{N,z}^2\tau_s^2}{1+\Omega_N^2\tau_s^2}
% \end{equation}
% and the averaging over the probability distribution function~\eqref{eq:dist_func} gives
% \begin{multline}\label{eq:S_z_lambda_0}
% 	\frac{\expval{S_z}}{S_0} = \frac{1}{3}\times \\ \times\left(1+\frac{4}{(\delta \tau_s)^2}-\frac{4\sqrt{\pi}}{(\delta \tau_s)^3}\exp[1/(\delta \tau_s)^2]\text{erfc}\qty[1/(\delta \tau_s)]\right),
% \end{multline}
% where $$\text{erfc}(r) = \frac{2}{\sqrt{\pi}}\int\limits_r^\infty\dd{t}e^{-t^2}.$$
% In the limit of the long spin relaxation time $ \tau_s\gg 1/\delta$, we obtain $\expval{S_z} = S_0/3$ similar to the analytical consideration~\eqref{eq:s0over3}. 

%\commentDima{We can provide the analytic expression for steady state $S_z$ (without averaging) either here or in the Appendix.}

\begin{figure}[t]
\centering
	\includegraphics[width=0.95\linewidth]{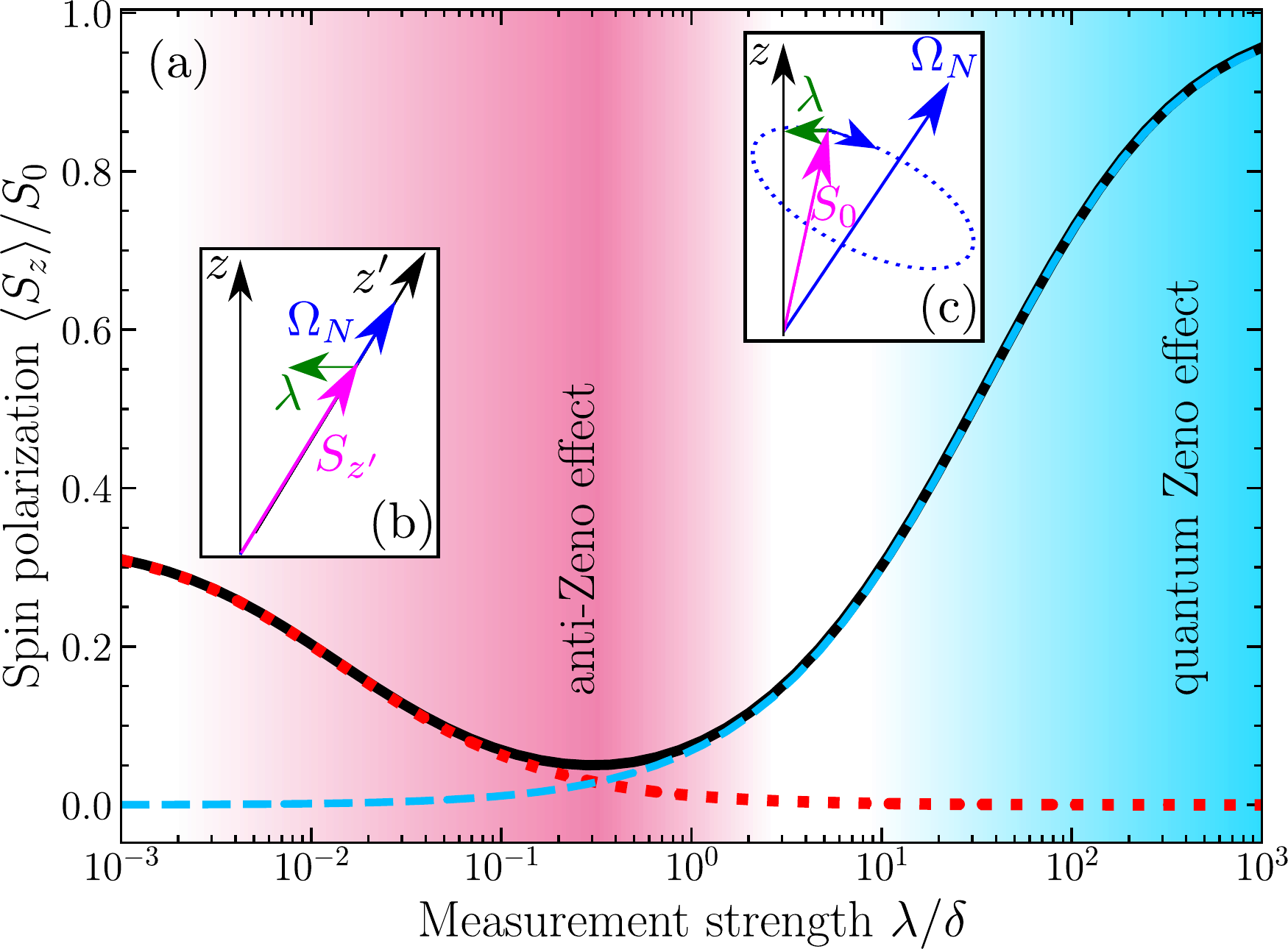} 
        \caption{(a) Steady state spin polarization as a function of the measurement strength for $\tau_s\delta=100$. The black solid curve is calculated numerically after Eqs.~\eqref{eq:Sz_no_average} and~\eqref{eq:dist_func}. The blue dashed and red dotted curves are calculated after Eqs.~\eqref{eq:az_an} and~\eqref{eq:z_an}, respectively. The background coral and light blue colors show the strengths of the quantum Zeno and anti-Zeno effects, respectively. (b) and (c) illustrate the orientation of electron spin (magenta arrow) with respect to the random nuclear field (blue arrow) in the quantum anti-Zeno and Zeno regimes, respectively.}
%       \commentDima{I suggest to change the labels according to the caption, move anti-Zeno inset to left and top a little bit, change blue dotted circle in it.} \commentDima{May be add light background similar to Fig.~\ref{fig:deph_time}?}} 
\label{fig:cont_meas}
\end{figure}

Figure~\ref{fig:cont_meas} shows the spin polarization $\expval{S_z}$ as a function of the measurement strength $\lambda$ (see Appendix~\ref{app:num} for the details on numerical averaging). One can see, that this dependence is nonmonotoneous: the spin polarization first decreases and then increases with increase of the measurement strength.

%\commentDima{Check that $\tau_s\delta\gg 1$ is stated.}

Qualitatively, this behaviour can be understood from Fig.~\ref{fig:deph_time}. First of all, in the absence of the spin measurement, $\lambda=0$, the spin polarization equals to $\braket{S_z}=S_0/3$, which corresponds to the long time limit in Fig.~\ref{fig:deph_time}, see Eq.~\eqref{eq:s0over3}. The measurement of $S_z$ projects the spin to the $z$ axis during the time $\sim1/\lambda$. If this time is short ($\lambda$ is large), this stabilizes the spin because of the initial quadratic decay in Fig.~\ref{fig:deph_time}. As a result, the nuclear induced spin relaxation becomes inefficient and the steady state spin polarization approaches $S_0$. This is the quantum Zeno effect.

Moreover, if the measurement time is of the order of $1/\delta$, the measurements project the spin to the $z$ axis at the minimum of the curve in Fig.~\ref{fig:deph_time}, which cancels the spin revival at longer times to $S_0/3$. As a result, the steady state spin polarization is much smaller than $S_0$. The acceleration of the nuclear induced spin relaxation is the manifestation of the quantum anti-Zeno effect.%\commentDima{Mention anti-Zeno in the introduction.}

%\addDima{Qualitatively, the two effect can be understood from Fig.~\ref{fig:deph_time}. The measurement of $S_z$ projects the spin to the $z$ axis during the time $\sim1/\lambda$. If this time is short ($\lambda$ is large) these stabilizes the spin because of the initial quadratic decay in Fig.~\ref{fig:deph_time}. By contrast, if the measurement time is of the order of $1/\delta$, the measurements project the spin to the $z$ axis at the minimum of the curve in Fig.~\ref{fig:deph_time}, which cancels the spin revival at longer times to $S_0/3$.}

%   One can see that if the measurement time 
Below we describe the two regimes in more detail and obtain the corresponding analytic expressions for the spin polarization.

% the behavior on the spin polarization non-monotonously depends on the measurement strength $\lambda$. This is the manifestation of the quantum Zeno and anti-Zeno effects. 

% The steady state solution of~\eqref{eq:spin_kin_res} and averaging over~\eqref{eq:dist_func} for the arbitrary  value of $\lambda$ could be done numerically. The details of the numerical derivation are discussed in Appendix~\ref{app:num}. The results of the calculation are shown in Figure~\ref{fig:cont_meas}.

% From the Figure~\ref{fig:cont_meas} one can see, that the behavior on the spin polarization non-monotonously depends on the measurement strength $\lambda$. This is the manifestation of the quantum Zeno and anti-Zeno effects.

\subsection{Quantum Zeno effect}

In the limit of $\lambda\gg\delta$ the measurement strongly suppresses the transverse spin components, as can be seen from Eq.~\eqref{eq:spin_kin_res}. As a result, the electron spin is almost parallel to the $z$ axis, as illustrated in Fig.~\ref{fig:cont_meas}(c).

  The transverse spin components appear in the first order in the random nuclear field and are given by
  \begin{equation}\label{eq:Zeno_Sy}
    S_x = \frac{\Omega_{N,y}S_z}{2\lambda},
    \quad
    S_y = -\frac{\Omega_{N,x}S_z}{2\lambda}.
  \end{equation}
Then in the second order we obtain from Eq.~\eqref{eq:spin_kin_res}
\begin{equation}
  \label{eq:Zeno_relaxation}
  \dot{S}_z(t)=g-\frac{S_z(t)}{\tau_s^{\text{eff}}},
\end{equation}
where the effective spin relaxation rate is~\cite{noise-trions}
\begin{equation}
  \label{eq:Zeno_eff}
  \frac{1}{\tau_s^{\text{eff}}} = \frac{1}{\tau_s}+\frac{\Omega_{N}^2\sin^2(\theta)}{2\lambda}.
\end{equation}

In the steady state the electron spin polarization equals to $g\tau_s^{\text{eff}}$ in agreement with Eq.~\eqref{eq:Sz_no_average}. Averaging it over the distribution function~\eqref{eq:dist_func} we obtain the average spin polarization in the form
\begin{equation}
  \label{eq:z_an}
  \frac{\expval{S_z}}{S_0} = -\nu \text{Ei}(-\nu)\exp(\nu),
\end{equation}
%\commentDima{I replaced $\text{Ei}(\nu)$ with $\text{Ei}(-\nu)$.}
 where $\nu = 2\lambda/(\tau_s\delta^2)$ and $\text{Ei}(x) = -\int\limits_{-x}^{\infty}\e^{-t}/t\dd{t}$
 is the exponential integral function. This expression is shown by the blue dashed curve in Fig.~\ref{fig:cont_meas}(a). One can see, that it agrees with the numerical calculations very well at ${\lambda/\delta > 1}$.

From Eq.~\eqref{eq:Zeno_eff} one can see that the measurement suppresses the nuclei induced spin relaxation due to the quantum Zeno effect. In particular, in the limit of $\lambda\to\infty$ one has $\tau_s^{\text{eff}}=\tau_s$, so the role of the hyperfine interaction is completely suppressed and the spin polarization $\expval{S_z}$ equals to $S_0$.

Thus the quantum Zeno effect can be used in realistic experimental conditions to increase the spin relaxation time. In the next subsection we describe the opposite quantum anti-Zeno effect, which allows one to accelerate the spin relaxation.

% In fact, if similar to the definition of $S_0$ we define $\expval{S_z} \equiv g \tau_s^{\text{eff}}$, where $\tau_s^{\text{eff}}$ has the meaning of  the effective spin relaxation time, we could rewrite~\eqref{eq:S_z_Zeno} as
% \begin{equation}
% 	\frac{1}{\tau_s^{\text{eff}}} = \frac{1}{\tau_s}+\NL{\frac{\Omega_{N}^2\sin^2(\theta)}{2\lambda }}.
% \end{equation}
% From this equation we see, that for $\lambda \gg \Omega_{N,x}\approx \delta$ the effective spin relaxation time $\tau_s^{\text{eff}} \to \tau_s$, while in the case $\lambda = 0$ we have from~\eqref{eq:s0over3} $\tau_s^{\text{eff}} = \tau_s/3$. That means, that measurement suppress the dynamics of the spin in the Overhauser field and the quantum Zeno effect~\cite{facchi_2008,Zeno_PRB} takes place. Our equations~\eqref{eq:Zeno_Sy} and~\eqref{eq:S_z_Zeno} show that it happens because large value of~$\lambda$ in~\eqref{eq:spin_kin_res} leads to the fast decay of $S_{x,y}$ components and does not give an opportunity to preccess in nuclear magnetic field \NL{(see the inset~(a) in Fig.\ref{fig:cont_meas})}. From the Figure~\ref{fig:cont_meas} one can see, that exact result of the numerical calculation smoothly crosses over two obtained analytical expressions.

\subsection{Quantum anti-Zeno effect}
In the limit of $\lambda\ll\delta$ we obtain from Eq.~\eqref{eq:Sz_no_average}
\begin{equation}
  S_{z} = \frac{S_0\cos^2(\theta)}{1+2\lambda\tau_s\sin^2(\theta)}.
\end{equation}
This expression can be averaged over the random nuclear fields analytically with the result
\begin{equation}
  \label{eq:az_an}
  \frac{\expval{S_z}}{S_0} = \frac{1}{2\lambda \tau_s}\qty[\sqrt{1+2\lambda \tau_s\over 2\lambda \tau_s}\text{arctanh}\qty(\sqrt{2\lambda \tau_s\over 1+2\lambda \tau_s})-1].
\end{equation}
This dependence of the spin polarization on the measurement strength is shown in Fig.~\ref{fig:cont_meas}(a) by the red dotted curve.

To qualitatively understand this regime, we note that the electron spin dynamics is dominated by the spin precession in the random nuclear field in this case. Therefore, the spin is almost parallel to the direction of $\bm\Omega_N$, which we denote as the $z'$ axis, see Fig.~\ref{fig:cont_meas}(b). From Eq.~\eqref{eq:spin_kin_res} we find that the dynamics of this component is described by
\begin{equation}
  \label{eq:anti-Zeno}
  \dot{S}_{z'} = - \frac{S_{z'}}{\tau_s}-2\lambda S_{z'} \sin^2(\theta)+g \cos(\theta).
\end{equation}
One can see that the spin generation along this axis is suppressed by the factor $\cos(\theta)$, while the measurement-induced relaxation of $S_x$ and $S_y$ contributes to the relaxation of $S_{z'}$ with the factor $\sin^2(\theta)$. This is because the measurement suppresses the transverse spin components, which leads to the small deviation of the average electron spin from the axis $z'$. Then the spin precession in the nuclear field leads to the additional spin relaxation. Acceleration of the spin relaxation, can be described by the effective spin relaxation rate
\begin{equation}
  \label{eq:Anti_eff}
  \frac{1}{\tau_s^{\text{eff}}}=\frac{1}{\tau_s}+2\lambda\sin^2(\theta),
\end{equation}
which clearly reveals the quantum anti-Zeno effect.

Eqs.~\eqref{eq:Zeno_eff} and~\eqref{eq:Anti_eff} show that the spin relaxation time decreases with decrease of $\lambda$ at $\lambda\gg\delta$ and decreases with increase of $\lambda$ at $\lambda\ll\delta$. So the minimum in the spin relaxation time and the steady state spin polarization unavoidably takes place at the crossover from the quantum Zeno to anti-Zeno regime, see Fig.~\ref{fig:cont_meas}(a). Moreover, one can see that the analytical limiting Eqs.~\eqref{eq:z_an} and~\eqref{eq:az_an} together describe the spin polarization very well in the whole range of
  $\lambda$. The crossover between the two curves takes at $\lambda\sim\delta/2$, when the quantum anti-Zeno effect is the strongest.
%
  % Comparing this expression with Eq.~\eqref{eq:Zeno_eff} for the quantum Zeno effect one can see that the anti-Zeno effect is the strongest at $\lambda\sim\delta/2$, which agrees with the minimum of the spin polarization shown in Fig.~\ref{fig:cont_meas}(a).}

%\NL{\sout{As usual, it take place when the measurement strength is smaller than the inverse time of the nonexponential dynamics, \NL{\sout{$\lambda\ll\delta$}} \NL{$\lambda \leq \delta$}}}. \NL{Comparing the $\lambda$ dependence of this  expression with the quantum Zeno effect case~\eqref{eq:Zeno_eff}, one can see, that the minimal effective relaxation rate would be achieved at the intersection of~\eqref{eq:Zeno_eff} and~\eqref{eq:Anti_eff}, namely at the value $\lambda \sim \delta/2$. This is exactly the coordinate of the curve extrema in Fig.~\ref{fig:cont_meas}.   }

%In the case of $\lambda\neq 0$, from the equation~\eqref{eq:az_an} we obtain $\expval{S_z}/S_0<1/3$. This happens because the term proportional to $\lambda$ in Eq.~\eqref{eq:spin_kin_res} acts like the additional to the spin dephasing on nuclei mechanism of the relaxation of $S_z$ component. Since, as discussed before, $\lambda$ represents measurement strength this effect is spin dynamic acceleration due to the measurements and can be called quantum anti-Zeno effect~~\cite{kaulakys_quantum_1997,facchi_2008,bednorz_nonclassical_2012}.

\section{Discussion}
\label{sec:Disc}
%The Quantum Zeno and anti-Zeno effects could be predicted only by analyzing the spin precession in the Overhauser field in the time picture. Assuming that spin relaxation time is larger than the period of the spin precession in the Overhauser field $\tau_s \gg 1/\delta$, then the spin dephasing with time is expressed via~\cite{book_Glazov}
%\begin{equation}
%	\bm S(t) = \frac{\bm S_0}{3}+\frac{2\bm S_0}{3}\qty(1-\frac{t^2\delta^2}{2})\exp(-\frac{t^2 \delta^2}{4}).
%\end{equation}
%Then, depending on the measurement strength $\lambda$ the time of the 

In this work we considered the continuous spin measurement. It can be always presented as a sequence of weak pulsed measurements with the short repetition period $T_R\ll 1/\lambda$. Generally, the quantum back action of each pulse can be described using the Krauss operator~\cite{facchi_2008,bednorz_nonclassical_2012, Zeno_PRB}
\begin{equation}
  K(s) = \qty(2\eta\over \pi)^{1/4}\e^{-\eta(s-\sigma_z/2)^2},
\end{equation}
where $\eta = 4\lambda T_R$ and $s$ is a possible outcome of the $S_z$ measurement. The measurement without postselection modifies the density matrix as~\cite{bednorz_nonclassical_2012}
\begin{equation}\label{eq:Kr_den}
  \rho^{\text{(after)}} = \int \dd{s} K(s)\rho^{\text{(before)}}K(s).
\end{equation}
This is equivalent to
  \begin{equation}
    \label{eq:eta}
    S_z^{\text{(after)}} = S_z^{\text{(before)}}, \quad S_{x,y}^{\text{(after)}} = \e^{-\eta/2} S_{x,y}^{\text{(before)}}.
  \end{equation}
In the same time, the spin measurement by resonant linearly polarized optical pulse modifies the spin as~\cite{PRC_General,yugova_pump-probe_2009}
\begin{equation}
  S_z^{\text{(after)}} = S_z^{\text{(before)}}, \quad S_{x,y}^{\text{(after)}} = (1-P) S_{x,y}^{\text{(before)}},
\end{equation}
%\commentDima{Can we describe elliptically polarized pulses?}
where $P$ is the probability of the trion excitation. From the comparison with Eq.~\eqref{eq:eta} one can see that $1-P = e^{-\eta/2}$. Under the above assumption we obtain
\begin{equation}
  \lambda = \frac{P}{2T_R},
\end{equation}
which establishes the relation between the power of the pulses, $P$, the repetition period, $T_R$, and the measurement strength $\lambda$. Thus our theory describes also the quantum Zeno and anti-Zeno effects under weak pulsed spin measurements.

%With this relation, the above results remain valid for $P\ll 1$.

% For weak measurements $\eta \ll 1$ it follows, that the equations~\eqref{eq:az_an} and~\eqref{eq:z_an} should work for the optical pulsed measurements with the substitution 
% \begin{equation}
% 	\lambda \to \frac{1-Q^2}{2T}.
% \end{equation}

Moreover, for the strong probe pulses, $P\lesssim 1$, our theory also qualitatively describes the Zeno effects.
%allows one to qualitatively understand the effect of the measurement back action.
In particular, one can consider the $\pi$ pulses, which are described by $P=1$. In this case, if the repetition period is short, $T_R\ll1/\delta$, then $\lambda\gg\delta$, and the quantum Zeno effect takes place. This is in agreement with the initial quadratic decay of $\expval{S_z}$, shown in Fig.~\ref{fig:deph_time}. By contrast, at $T_R\sim1/\delta$ one has $\lambda\sim\delta$, so the quantum anti-Zeno effect takes place. Indeed, in this case the probe pulses project the spin to the $z$ axis in the minimum of the curve in Fig.~\ref{fig:deph_time}, which accelerates the spin relaxation.

% We could expect, that our theory is also valid for the projective measurements when $\eta \gg 1$.  From Fig.~\ref{fig:cont_meas} one see, that the Zeno effect is realized when $\lambda/\delta \gg 1$. So, from the condition on $\eta$ we obtain, that this is similar to the pulsed measurements with small period $T$. The quantum Zeno effect in this limit is actually expected, since projective pulsed measurements with the period $T<1/\delta$  will froze spin dynamics in the state where spin dephasing has not happened yet (see Fig.~\ref{fig:deph_time}).

% On the other hand, for the same reason, the anti-Zeno effect should be realized for the measurement with $T>1/\delta$, but still $T\ll \tau_s$. It is true, because on such times the dephasing to the spin polarization value $S_0/3$ happens and measurements will work as an additional relaxation mechanism decreasing spin polarization. The indication bar in Fig.~\ref{fig:deph_time} shows the typical periods to achieve quantum Zeno and anti-Zeno effects in the case of the strong measurements.

\section{Conclusion}
\label{sec:Concl}

To conclude, we have described the effect of the quantum measurement back action under continuous measurement and orientation of electron spins in quantum dots by elliptically polarized light. We have demonstrated that the non-Markovian electron spin dynamics driven by the hyperfine interaction with the host lattice nuclear spins allows for both quantum Zeno and anti-Zeno effects. For the large light power the nuclear induced spin relaxation is suppressed, which leads to the increase of the steady state spin polarization due to the quantum Zeno effect. For moderate power of light the nuclear induced spin relaxation is accelerated, which results in the suppression of the spin polarization due to the quantum anti-Zeno effect. The theoretical predictions can be directly compared to the future experimental results for ensembles of self-assembled charged quantum dots.

%analyzed theoretically the spin polarization in the ensemble of charged quantum dots under the weak measurement by light and predicted the manifestation of quantum Zeno and anti-Zeno effects. We provide the physical interpretation for those effects in terms of spin motion and obtained analytical expressions for the spin polarization~\eqref{eq:az_an} and~\eqref{eq:z_an} in both cases.

\acknowledgments

We thank \href{https://orcid.org/0000-0001-7813-682X}{A. Greilich}, V. Nedelea and \href{https://orcid.org/0000-0002-5315-8720}{E. Evers} for showing their preliminary but encouraging experimental results, RF President Grant No. MK-5158.2021.1.2, and the Foundation for the Advancement of Theoretical Physics and Mathematics "BASIS". The derivation of the analytical expressions and the numerical calculations were supported by the Russian Science Foundation Grant No. 21-72-10035.

\appendix

\section{Numerical details}\label{app:num}

To average the dimensionless quantity $\mathcal{S}(\Omega_N\tau_s,\theta)=S_z/S_0$ over the distribution function~\eqref{eq:dist_func} we use the Gauss-Laguerre quadrature to speed up the calculations. Then the average can be calculated as
\begin{multline}
  \expval{\mathcal{S}}= \frac{1}{\sqrt{\pi}}\int \dd{y} e^{-y}\sqrt{y}\int \dd{\theta}\sin(\theta)\mathcal{S}(\sqrt{y}\tau_s\delta,\theta)  \\ \approx\frac{2}{\sqrt{\pi}}\sum_{i = 1}^N w_i  \overline{\mathcal{S}}(y_i\tau_s\delta)\sqrt{y_i},
\end{multline}
where $y = \Omega_N^2/\delta^2$, 
\begin{equation}\label{eq:Append_av}
\overline{f} = \frac{1}{2}\int f\sin(\theta)\dd\theta,
\end{equation}
$w_i$ and $y_i$ are the weights and the roots according to the Gauss-Laguerre quadrature scheme~\cite{salzer_table_1949}. The averaging in Eq.~\eqref{eq:Append_av} is performed using the Simpson's rule. In the calculation we use $N  = 15$ and check that larger $N$ yields the same results.

% In the numerical calculation we average the Eq.~\eqref{eq:Sz_no_average} and find dimensionless $\mathcal{S} \equiv \frac{S_z}{S_0}$ as a function of \NL{$(\Omega_N \tau_s,\theta)$}. Then we average it over the function~\eqref{eq:dist_func} using Gauss-Laguerre quadrature to speed up the calculations
% \begin{multline}
% 	\expval{\mathcal{S}}=\frac{1}{(\sqrt{\pi}\delta)^3}\int \dd{\bm \Omega}_N  \mathcal{S}(\Omega_N \tau_s,\theta)e^{-\Omega_N^2/\delta^2} =\\ = \frac{2\pi}{2(\sqrt{\pi})^3}\int \dd{y} e^{-y}\sqrt{y}\int \dd{\theta}\sin \theta  \mathcal{S}(\sqrt{y}\tau_s\delta,\theta)  = \\ =\frac{1}{\sqrt{\pi}}\sum_{i = 1}^N w_i  \overline{\mathcal{S}}(\tau_s\delta y_i)\sqrt{y_i},
% \end{multline}
% where we introduced $y = \Omega_N^2/\delta^2$ and  
% \begin{equation}\label{eq:Append_av}
% \overline{f} = \int \dd{\theta}\sin\theta  f,
% \end{equation}
% $w_i$  and $y_i$ weights and roots according to the Gauss-Laguerre quadrature scheme~\cite{salzer_table_1949}. The angle averaging~\eqref{eq:Append_av} is done using Simpson's rule. In the calculations we use $N  = 15$ and check that futher increase of $N$ does not affect the results.

%\newpage~\newpage

\bibliography{Zeno_CW_ds2}

\end{document}